\documentclass[12pt]{article}
\usepackage{graphicx}

\def\pd{\partial}
\def\a{\alpha}
\def\b{\beta}
\def\dl{\delta}
\def\s{\sigma}

\def\lam{\lambda}
\def\Lam{\Lambda}
\def\hg{{\hat g}}
\def\hnabla{{\hat \nabla}}
\def\hR{{\hat R}}
\def\hDelta{\hat{\Delta}}

\def\gm{\gamma}

\def\om{\omega}

\def\bfx{{\bf x}}
\def\bfr{{\bf r}}
\def\bfk{{\bf k}}
\def\bfp{{\bf p}}
\def\bfn{{\bf n}}
\def\QG{{\rm QG}}
\def\P{{\rm P}}
\def\D{{\rm D}}
\def\G{{\rm G}}
\def\cl{{\rm cl}}
\def\sq{\sqrt}
\def\e{\hbox{\large \it e}}

\def\fr{\frac}
\def\st{\stackrel}
\def\pp{\prime}

\def\bb{\begin{equation}}
\def\ee{\end{equation}}
\def\bba{\begin{eqnarray}}
\def\eea{\end{eqnarray}}

\begin{document}

\begin{titlepage}

\begin{tabbing}
   qqqqqqqqqqqqqqqqqqqqqqqqqqqqqqqqqqqqqqqqqqqqqq
   \= qqqqqqqqqqqqq  \kill
         \>  {\sc KEK-TH-934}    \\

\end{tabbing}

\vspace{1.5cm}

\begin{center}
{\Large {\bf CMB Anisotropies Reveal Quantized Gravity} }
\end{center}

\begin{center}
{Ken-ji Hamada$^1$ and Tetsuyuki Yukawa$^2$} 
\end{center}

\begin{center}
{}$^1${\it Institute of Particle and Nuclear Studies, KEK, Tsukuba 305-0801, Japan}
{}$^2${\it Coordination Center for Research and Education, \\ 
The Graduate University for Advanced Studies (Sokendai), Hayama 240-0193, Japan}
\end{center}

\begin{abstract}
A novel primordial spectrum with a dynamical scale of quantum gravity origin is 
proposed to explain the sharp fall off of the angular power spectra at low multipoles 
in the COBE and WMAP observations. 
The spectrum is derived from quantum fluctuations of the scalar curvature in 
a renormalizable model of induced gravity. 
This model describes the very early universe by the conformal field fluctuating about an inflationary background with the expansion time constant of order of the Planck mass.
 
\vspace{5mm}

\end{abstract}
\end{titlepage}

The discovery of anisotropies in the cosmic microwave background (CMB) by
the Cosmic Background Explorer (COBE)~\cite{cobe} 
and the Wilkinson Microwave Anisotropy Probe (WMAP)~\cite{wmap} 
has opened a new frontier on spacetime physics. 
Amazingly, if we believe the idea of inflation~\cite{guth, starobinsky,hamada01} 
originally introduced to resolve the flatness and horizon problems, 
the observed CMB anisotropies provide us information about dynamics beyond the Planck scale. 
Thus, we are now on the stage of revealing and verifying the quantum aspect of spacetime.

When we take a look on the angular power spectra, the sharp fall off of the $l=2$, and $3$ multipole components from the nearly constant behavior up about to $l=40$ is apparent. 
The constant behavior in low multipole components has been expected 
as the Sachs-Wolfe effect~\cite{sw} 
assuming the Harrison-Zel'dovich spectrum~\cite{hz} 
for the initial fluctuation. 
This deviation has been regarded just a statistical fluctuation known as cosmic variance, i.e. a specific case among ensemble of universes~\cite{aw}.
We shall not put the deviation into statistics, but look for the cause in dynamics, regarding it as a reflection of the new physical scale possessing in the quantum theory of spacetime.

Despite the common belief that there exists no consistent quantum theory of gravity at hand, we propose a model of quantum gravity which explains the sharp fall of the spectrum, and also ignites the inflation naturally without any additional fields.
The model we employ is a renormalizable model 
based on the conformal gravity in 4-dimension~\cite{hamada02}.
This model seems to be in the same universality class 
as the 4-dimensinal simplicial quantum gravity~\cite{hey}.

By this model the evolution of universe proceeds as follows: just after the birth there is no spacetime, but quantum fluctuation, which is dominated by the conformal mode.
Physically, such a state can be imaged by the 4-dimensional simplicial manifold with varying connectivity of triangulated space, i.e. so called dynamical triangulation. 
Soon, the conformal symmetry is slightly broken to develop the classical solution inflating exponentially with a time scale 
of the Planck mass~\cite{hamada01}. 
During the inflation, large-scale patterns of quantum fluctuation is preserved, 
for example, in the two point correlation with super-horizon separations. 
The inflational expansion will eventually terminate to the big bang 
by the effect of the traceless mode of gravity.

\paragraph{The Model}
The renormalizable model of quantum gravity is defined 
by the action~\cite{hamada02},\footnote{
The action is determined to be in this form by requiring the integrability 
and the asymptotic freedom. 
}
\bb
   I=\int d^4 x \sq{-g} \left\{ 
     -\fr{1}{t^2}C_{\mu\nu\lam\s}^2 -bG_4 + \fr{M_\P^2}{2} R 
     -\Lam_{\rm COS} -\fr{1}{4} tr \left( F_{\mu\nu}^2 \right) + \cdots \right\},   
            \label{action}
\ee
where $M_\P =1/\sq{8\pi G}$ is the reduced Planck mass, and $\Lam_{\rm COS}$ is the cosmological constant. $C_{\mu\nu\lam\s}$ is the Weyl tensor, and $G_4$ is the Euler density. 
The dots denote conformally coupled scalar and fermionic matter fields in addition 
to vector fields $F_{\mu\nu} $.

The metric field is decomposed to the conformal mode $\phi$ 
and the traceless mode $h^{\lam}_{~\nu}$ as
\bb
      g_{\mu\nu} = \e^{2\phi} \hg_{\mu\lam} 
               (\dl^{\lam}_{~\nu} + t  h^{\lam}_{~\nu} +\cdots ),   
\ee
with $tr(h)=0$. 
The traceless mode will be handled perturbatively in terms of the coupling $t$ on the background metric $\hg_{\mu\nu}$, while the conformal mode is treated {\it exactly}. 
The beta function of the renormalized coupling $t_r$ for the traceless mode has been calculated within the lowest order, which indicates the asymptotic freedom: $\b_t =-\b_0 t_r^3$ with  
$\b_0= \{ ( N_{\rm X} +3N_{\rm W} +12N_{\rm A} )/240 + 197/60 \}/(4\pi)^2$,
where $N_{\rm X}$, $N_{\rm W}$ and $N_{\rm A}$ are the numbers of scalar fields, Weyl fermions and gauge fields, respectively.  
The asymptotic freedom implies at very high energies above the Planck mass 
the Weyl tensor should vanish, and spacetime fluctuations are dominated 
by the conformal field.  
Physical states in this phase do not look like ordinary point-particle states, but composite states classified by representations of the conformal algebra~\cite{hh}.  

The partition function of the model, describing states in the very early universe, is given by
\bb
   Z \vert_{t=0}  
   = \int [d\phi dh dA \cdots]_\hg 
                   \exp \left( i I_{\rm CFT} \right).
           \label{CFT}
\ee 
Here, we rewrite the diffeomorphism invariant measure defined on the metric $g_{\mu\nu}$ to the practical measure defined on the background metric $\hg_{\mu\nu}$~\cite{kpz}. 
Consequently the kinetic term of $\phi$~\cite{riegert,am} appears in the action in order to preserve diffeomorphism invariance. Dynamics for the conformal mode is then effectively described by the conformal field theory (CFT$_4$) with the action
\bb
  I_{\rm CFT} = -\fr{b_1}{(4\pi)^2} \int d^4 x \sq{-\hg} 
            \left\{ 2\phi \hDelta_4 \phi 
            + \left({\hat G}_4 -\fr{2}{3}\hnabla^2 \hR \right) \phi \right\} 
           + I\vert_{t=0},  
\ee
where $\sq{-\hg}\hDelta_4 = \sq{-\hg}(\hnabla^4 + \cdots)$ is the conformally 
invariant 4-th order adjoint operator.
Here, the hat on each character indicates to use $\hg_{\mu\nu}$ instead 
of $g_{\mu\lam}$ for the definition.

The coefficient $b_1$ has been calculated within the lowest order in $t$ to be
\bb
    b_1 = \fr{1}{360}\left( N_{\rm X} +\fr{11}{2}N_{\rm W} +62 N_{\rm A} \right) 
          +\fr{769}{180}.
\ee
At higher orders~\cite{hamada02}, interaction terms like 
$\phi^{n+1}\hDelta_4 \phi$, $\phi^n C_{\mu\nu\lam\s}^2$ and $\phi^n tr( F_{\mu\nu}^2 )$  
with $n \geq 1$ will be induced from the invariant measure in addition to the self-couplings 
of the traceless mode.

Dynamics of the traceless mode introduces the scale parameter $\Lam_\QG$ in the similar manner as the running coupling of gauge theories,
\bb
   \a_\G =\fr{t_r^2(p)}{4\pi}= \fr{1}{4\pi \b_0}\fr{1}{\log (p^2/\Lam_\QG^2)}, 
         \label{alphaG}
\ee  
where $\Lam_\QG=\mu \e^{-1/2\b_0 t_r^2(\mu)}$, and $\mu$ is a renormalization mass scale.
The appearance of $\Lam_\QG$ will be shown to influence low multipole components of the angular power spectra.

\paragraph{Conformal Gravity Scenario of Inflation}
There are three mass scales involved in our model, namely the Planck mass $M_\P$, 
the dynamical scale $\Lam_\QG$, and the cosmological constant $\Lam_{\rm COS}$. 
We set their order~\footnote{
The condition $M_\P \gg \Lam_\QG$ implies that quantum effects turn on at much larger scale than the Planck length so that not only the spacetime singularity but also the horizon disappear at the final stage of the black hole evaporation.
}
as
\bb
   M_\P \gg \Lam_\QG \gg \Lam_{\rm COS}. 
   \label{scale-relation}
\ee
In the very early epoch when the space extends much smaller than the Planck scale typical energy is higher than $M_\P$. Then, corrections of order $\a_\G$ can be neglected and the dynamics is governed by the CFT$_4$ with the full conformal invariance. 
There exists no classical spacetime and the universe is filled with quantum fluctuation of the conformal field.

As the space expands the energy gets lowered to $M_\P$, and the Einstein action becomes effective. 
The conformal field fluctuates around a solution $\phi_\cl$ of the classical equation,   
\bb
    -\fr{b_1}{(4\pi)^2}  4\pd_\eta^4 \phi_\cl   
       +6 M_\P^2 \e^{2\phi_\cl} \left\{ \pd_\eta^2 \phi_\cl 
                                         + (\pd_\eta \phi_\cl)^2 \right\} 
    =0,
         \label{eom}
\ee
where we look for the homogeneous solution $\phi_\cl =\phi_\cl (\eta)$, depending only on the conformal time $\eta$.

Introducing the proper time $\tau$ defined by $d\tau = a(\eta) d\eta$, 
and writing  $a=e^{\phi_\cl}$, the equation for $H=\dot{a}(\tau)/a(\tau)$, 
where dot represents the derivative in terms of the proper time, reads~\cite{hamada01} 
\bb
   b_1 \Bigl( \st{...}{H} +7 H \st{..}{H} +4 \st{.}{H^2} 
              +18H^2\st{.}{H} +6H^4 \Bigr)
    - 24\pi^2 M_\P^2 \left( \st{.}{H} +2H^2 \right) =0. 
\ee 
It has two stationary solutions: a {\it stable} inflationary solution 
with $H=H_\D$, 
\bb
      a(\tau) \propto \e^{H_\D \tau}, \qquad  H_\D=\sq{\fr{8\pi^2}{b_1}}M_\P, 
        \label{desitter}
\ee
and an unstable solution with $H=0$.
Starting with the unstable solution, the solution immediately shifts to the stable 
one (\ref{desitter}) by an infinitely small perturbation $H \approx \epsilon$ (Fig.1). 
Quantum fluctuation of the conformal mode still keeps a long range correlation over the horizon scale on the de Sitter background (\ref{desitter}).
\begin{figure}
\begin{center}
\begin{picture}(300,200)(80,0)
\put(50,0){\includegraphics[width=13cm,clip]{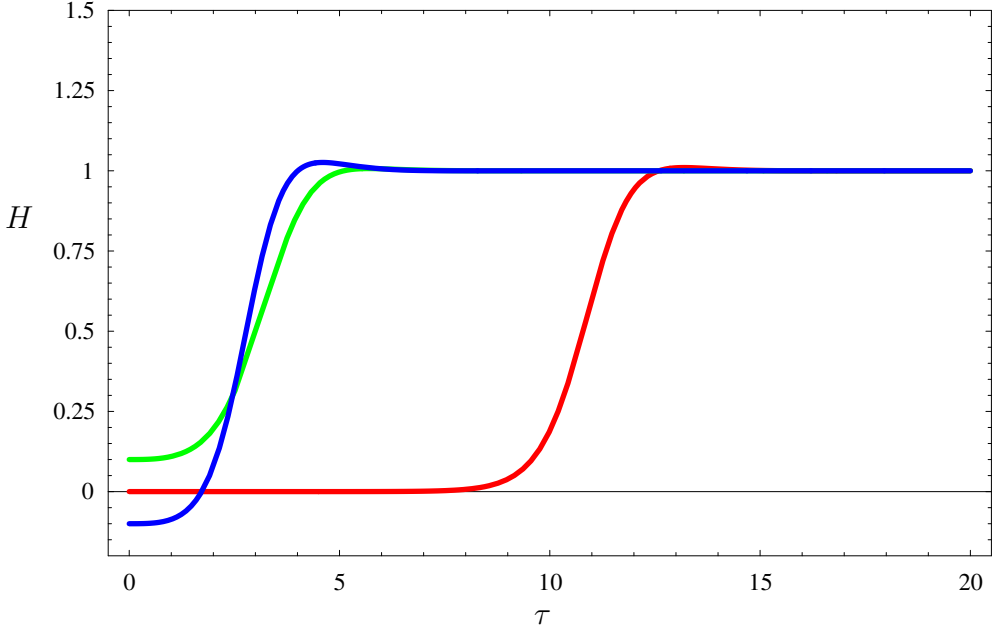}}
\put(40,140){$H$}
\put(240,-10){$\tau$}
\end{picture}
\end{center}
\caption
{\label{Fig.1}
The time evolution of $H$ from near the unstable solution. $H_\D$ is normalized to $1$. 
$H(0)$ is taken to be $0.1$(green),$0.0001$(red), and $-0.1$(blue).
}
\end{figure}

The inflation will terminate eventually at the energy scale $\Lam_\QG$, where the effective coupling strength to the traceless mode diverges (\ref{alphaG}), and the action will depart from $I_{\rm CFT}$ considerably. 
At this stage we expect the correlation of field fluctuations becomes short range and spacetime becomes classical. 
Because of the induced higher-order interactions, the $b_1 \pd_\eta^4 \phi$ term in the equation of motion (\ref{eom}) would change to $b(\phi) \pd_\eta^4 \phi$, where $b(\phi)$ is expected to vanish as $\phi \rightarrow \infty$ or $\a_\G \rightarrow \infty$, so that the Einstein gravity dominates in the classical equation. 
The universe would make a sharp transition from the inflation era to the Friedmann era. 
The primordial fluctuation we observe in the CMB is expected to be quantum fluctuation of the conformal mode developed right before this transition.

Since the inflation starts at the Planck scale $\tau_\P=1/H_\D$ and 
ends at $\tau_{\Lam}=1/\Lam_\QG$, the number of e-foldings of inflation is estimated as 
\bb
     {\cal N}_e = \log \fr{a(\tau_{\Lam})}{a(\tau_\P)} 
         =H_\D \left( \tau_{\Lam} -\tau_\P \right) \simeq  \fr{H_\D}{\Lam_\QG}.
\ee
Once the value of $\Lam_\QG$ is chosen to fit the observed data, we can check whether the model can give the right number of e-foldings large enough to solve the flatness and horizon problems.

\paragraph{CMB Anisotropies} 
The observed CMB anisotropies are considered to be a reflection of the gravitational potential on the last scattering surface where photons decouple to matter at the recombination period. 
The temperature fluctuation follows the relationship known as the Sachs-Wolfe 
effect~\cite{sw}, $\dl T/T =\Phi(\bfx_{\rm lss})/3$, 
where $\Phi(\bfx_{\rm lss})$ is the gravitational potential on the last scattering 
surface $\bfx_{\rm lss}$ in the comoving frame. 
The anisotropy may be expanded by the spherical harmonics,
$\dl T/T=\sum_{l,m} a_{lm}Y_{lm}$.

The angular power spectrum is obtained from the two point correlation function 
of the statistical ensemble. 
Since we observe only one universe and one sky, the statistics cannot be taken over ensemble of universes. Instead we take averages over ensemble of sub-systems made out of the universe. 
Thus, linearly independent members of the ensemble are counted at most $2l+1$, and we have the unavoidable limitation of the measurement precision with the statistical error $1/\sq{2 l+1}$, known as cosmic variance~\cite{aw}.
Assuming the ergodicity, the ensemble of sub-systems is considered to be Gaussian 
with $\langle a_{lm} \rangle =0$, 
and $\langle a_{lm } a_{l'm' }^* \rangle =C_l \dl_{ll'} \dl_{mm'}$. 
The two point correlation function is then expressed in terms of the angular 
power spectrum $C_l=\langle |a_{lm }|^2 \rangle $.

The ergodicity assumption is based on the mixing property of dynamics in the universe. 
When two points in space are located within the event horizon it is reasonable to assume the mixing property since matter, radiation and geometry are considered to couple strongly. 
However, for a pair of points in space with super-horizon separation there exists no dynamical correlation, and there is no reason to assume the statistical isotropy. 
In such a case we expect that the initial fluctuation property will be preserved in the inflating spacetime.

In order to relate the observed anisotropies to the quantum fluctuation, we consider the process into two steps: the step at the recombination time $\tau_{\rm rec}$, and the step at the dynamical transition $\tau_\Lam$.
At $\tau_{\rm rec}$ the potential fluctuation $\Phi$ is related to the density contrast, $\dl_\rho =\dl\rho/\rho$, through the Poisson equation, $ \vec{\nabla}^2 \Phi = 4\pi G \delta \rho $, and the Friedmann equation, $H^2=8\pi G \rho/3$, 
resulting $\dl_\rho (\bfk) = -(2/3) ( k/m_{\rm rec})^2 \Phi (\bfk)$, 
where $\bfk$ is the comoving wave number, and $m_{\rm rec}=aH$ evaluated at $\tau_{\rm rec}$. 
At $\tau_\Lam$ we assume the primordial density fluctuation precisely reflects the scalar curvature fluctuation, which is the unique observable with the general covariance:
\bb
    \fr{\dl R}{R} \biggr|_{\tau_\Lam^-} ~\sim~~ 
    \fr{\dl\rho}{\rho} \biggr|_{\tau_\Lam^+},
\ee
where $\tau_\Lam^{-(+)}$ is the time just before(after) $\tau_\Lam$. This density contrast at the large scale with super-horizon separations will be preserved until $\tau_{\rm rec}$.

The two point correlation function is defined by the quantum mechanical expectation of the scalar curvature contrast operator, $\dl_R = ( \dl R/R )_{op}$, as
\bb
  c_2(\theta)
   = \int \fr{d^3 \bfk}{(2\pi)^3} \fr{1}{4} \left( \fr{m_{\rm rec}}{k} \right)^4 
     \langle\langle 
      \dl_R (\bfk) \dl_R (-\bfk) 
      \rangle\rangle  
       e^{i\bfk \cdot (\bfn-\bfn') x_{\rm lss}}, 
\ee
where $\langle\langle ~ \rangle\rangle$ indicates the quantum mechanical expectation value, 
and $x_{\rm lss} =|\bfx_{\rm lss}|$.
Expanding exponential functions in terms of the spherical harmonic functions, 
and integrating over angles we obtain
the angular power spectrum as
\bb
      C_l 
      = \fr{1}{2\pi} \int \fr{dk}{k} \fr{m^4_{\rm rec}}{k} 
        \langle \langle  
      \dl_R (\bfk) \dl_R (-\bfk) 
        \rangle\rangle j_l^2 (kx_{\rm lss}), 
        \label{Cl}
\ee
where $j_l (x)$ is the spherical Bessel function.

The two point correlation function of the scalar curvature contrast $\dl R/R$ 
in CFT$_4$ is given by  
\bb
    \langle\langle  
      \dl_R (\tau_\Lam, \bfr) \dl_R (\tau_\Lam, \bfr^\pp) 
        \rangle\rangle 
     \sim \left( H_\D |\bfr -\bfr^\pp | \right)^{-2\Delta_R},
\ee
where $|\bfr-\bfr^\pp| =a(\tau_\Lam) |\bfx-\bfx^\pp| $ is the physical distance 
on the hypersurface at $\tau_\Lam$, and $\Delta_R $ is the scaling dimension of $\dl_R$.

In order to obtain the scaling dimension~\cite{kpz,amm}, 
let us illustrate the properties of operators in CFT$_4$. 
Any physical operator should be conformal invariant, and in this sense the dynamical fields, such as $\phi$, $h^{\mu}_{~\nu}$, $A_{\mu}$, are themselves not physical operators. 
A physical operator with the conformal dimension $\Delta$ is written by a composite field operator as $O_{\Delta}=\sq{-\hg}\e^{\gm\phi}{\cal O}_{\Delta_0}$, 
where $\gm$ is the conformal charge, and $\Delta_0$ is the naive dimension of 
the operator ${\cal O}_{\Delta_0}(\pd \phi, h, A, \cdots)$.  
Within the lowest order of $\a_\G$, the conformal charge is computed in CFT$_4$ 
to be $ \gm =2b_1 \{ 1- \sq{1-(4-\Delta_0)/b_1} \}$~\cite{am}.
For example, the cosmological constant and the scalar curvature operators, which have naive dimension 0 and 2, respectively, are written as $\sq{-\hg}\e^{\gm_0\phi}$ 
and $\sq{-\hg} \e^{\gm_2\phi} {\cal R} (\pd \phi)$.    
The curvature contrast operator is then given by $\dl_R=\e^{\gm_2\phi}{\cal R}/12H_\D^2$, 
where the denominator is the curvature of the de Sitter background (\ref{desitter}).

The scaling dimension of the scalar curvature $\Delta_R$ is determined from the scale transformation properties of operators: 
the operator with the scaling dimension $\Delta$ is considered to transform 
as $d^4x O_\Delta \rightarrow \om^{4-\Delta}d^4 x O_{\Delta}$ 
under the constant Weyl rescaling defined such that the cosmological constant operator 
transforms as $\Delta=\Delta_0=0$. The Weyl rescaling is equivalent to the constant 
shift $\phi \rightarrow \phi +(4/\gm_0)\log \om$.
By this shift the operator with the scaling dimension $\Delta$, $d^4 x O_\Delta$, 
changes to $\om^{4\gm/\gm_0} d^4 x O_\Delta$. 
Thus, we obtain the relation $ 4-\Delta = 4 \gm/\gm_0$, 
which gives $ \Delta_R = 4-4 \gm_2/\gm_0$ for the scalar curvature operator. 
The difference $\Delta-\Delta_0$ is the anomalous dimension of the operator.

The Fourier component of scalar curvatures correlation function is written as 
\bb
d^3 \bfp \fr{A^\pp}{H_\D^3} \left( \fr{|\bfp|}{H_\D} \right)^{2\Delta_R-3} 
   = d^3 \bfk \fr{A^\pp}{m_\lambda^3} \left( \fr{|\bfk|}{m_\lambda} \right)^{2\Delta_R-3},     
\ee
where $\bfp=\bfk/a(\tau_{\Lam})$ is the physical wave vector at $ \tau_{\Lam} $, 
and $\bfk$ is the comoving wave vector.  $A^\pp$ is a dimensionless normalization  constant 
and the constant $m_\lambda$ is given by $m_\lambda=a(\tau_\Lambda )H_\D$. 
The exponent $n=2\Delta_R-3$~\cite{amm} 
is the spectral index.  
The deviation from the Harrison-Zel'dovich spectrum~\cite{hz}, 
$n-1=2/b_1 + 4/b_1^2 + o\left( 1/b_1^3 \right)$, is the contribution from CFT$_4$.

In general, the conformal charges depend on the traceless mode, and thus the scaling dimension has correction from the traceless-mode coupling. 
Hence, the index of the correlation should be replaced by
\bb
      \bar{n}  
       =n +u \a_\G, 
       \label{index} 
\ee 
where $u$ is a positive constant.\footnote{
This constant is a calculable quantity, but is not given here, because renormalizations of composite operators with conformal charges are rather complicated. 
In practice the precise value does not matter much in the following discussion.
}

The two point correlation function involves integration over the momentum, which runs in the region where the coupling with the traceless mode is not negligible. In order to take into account the non-perturbative effect we replace the effective coupling constant $\a_\G$ 
by the running coupling (\ref{alphaG}).
We then obtain the angular power spectrum for large angles to be
\bb
   C_l 
    = \int^{\infty}_{\lam} \fr{dk}{k}j_l^2(kx_{\rm lss})P (k) 
\ee
with 
\bb
     P(k) = A \left( \fr{k}{m_\lambda} \right)^{n-1+\fr{v}{\log(k^2/\lam^2)}},
\ee
where $A=(A^\pp/2\pi) ( m_{\rm rec}/m_\lambda )^4$ 
and $v=u/4\pi\b_0$. 
$\lam$ is the comoving dynamical scale at the time $\tau_{\Lam}$ defined by 
$\lam =a(\tau_{\Lam}) \Lam_\QG$.

\paragraph{Spectrum}
We now determine parameters by comparing low multipole components of the angular power spectra from the WMAP observation. 
We focus on the sharp damping at $l=2$ and $3$ multipole components as an appearance of the dynamical scale of quantum gravity. 
This determines a value for the comoving dynamical scale to be about $\lam=3/x_{\rm lss}$. 
If we take $x_{\rm lss}=14000$ Mpc, we obtain $\lam=0.0002$ Mpc$^{-1}$. 
The number of e-foldings, ${\cal N}_e \simeq H_\D/\Lam_\QG = m_\lambda/\lam$, can be taken an arbitrarily large value according to the value $m_\lambda$. 
For example, we can choose $m_\lambda=0.02$ Mpc$^{-1}$ so that ${\cal N}_e=100$ which is large enough to solve the flatness problem. 
Since $H_\D$ is order of $10^{19}$ GeV, we have $\Lam_\QG \sim 10^{17}$ GeV.

The spectral index $n$ of the Standard Model ($n=1.41$ from $N_A=12$, $N_W=45$) predicts large blue spectrum. From the observation it looks favorable to make $n$ a little smaller by adding extra matter and gauge fields as the GUT or SUSY models suggest. The integrated Sachs-Wolfe(ISW) effect may shift the index up~\cite{hu-sugiyama}. 
However, the conclusive statement should be postponed until the overall analysis based on our spectrum is completed.
The CMB spectra, $l(l+1)C_l$, are calculated up to $l = 40$ for $n=1.1$, $1.2$, $1.3$ as shown in Fig.2 in company with the WMAP data.   
The normalization constant $A$ is chosen appropriately so that $l=6$ multipole components coincide with the observed value, and $v$ is taken to be $0.001$.
\begin{figure}
\begin{center}
\begin{picture}(300,200)(80,0)
\put(50,0){\includegraphics[width=13cm,clip]{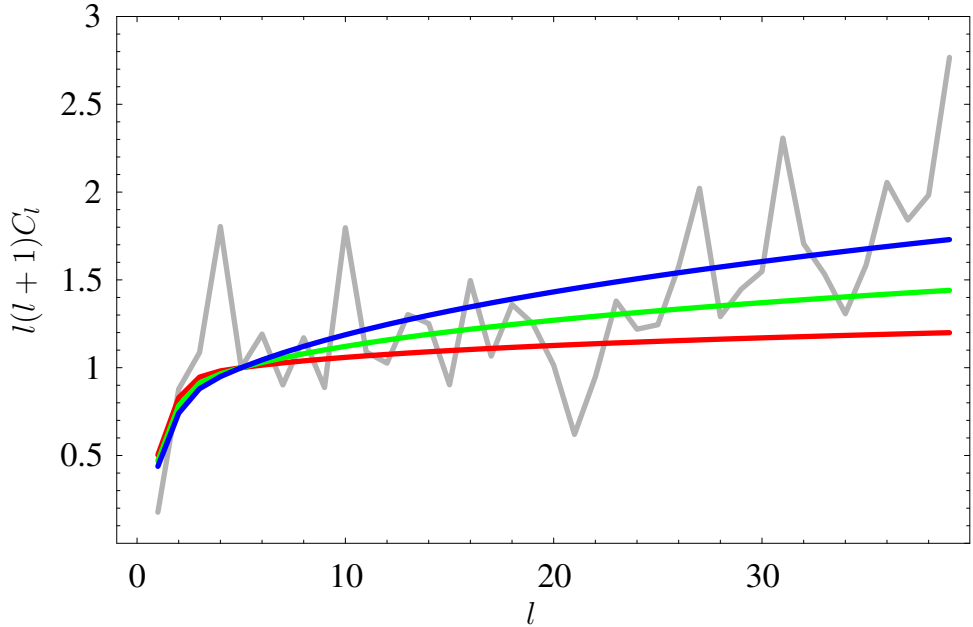}}
\put(45,100){\rotatebox{90}{$l(l+1)C_l$}}
\put(240,-10){$l$}
\end{picture}
\end{center}
\caption
{\label{Fig.2}
The CMB spectra for $n=1.1$(red), and $1.2$(green), $1.3$(blue) together with the WMAP data denoted by the zigzag line. 
}
\end{figure}

In this note we consider the angular spectrum only in the super-horizon region where the quantum nature of gravity shows up significantly. 
In order to achieve the overall fit to the data with our primordial spectrum, more detail consideration is necessary~\cite{hu-sugiyama,ll}.
For example, we have assumed that the unique decoupling time $\tau_\Lam$ for entire momentum range. 
However, if there is a time lag in the phase transition, the short scale delay will 
change $m_\lambda$ to be an increasing function of the comoving wave number $k$.

In summary, we have shown an inflationary scenario induced by quantum gravity and derived the CMB angular power spectrum at large angles.  
The sharp damping at low multipole components is interpreted to reflect the dynamical scale of quantum gravity. 
Since the primordial spectrum is produced by CFT$_4$, we expect that the tensor-to-scalar ratio is negligible and multi point correlation functions of non-Gaussian type exist.

We would like to add our big bang scenario for completeness. 
The universe is, initially, a pure quantum state of the conformal field with a long-range correlation. 
As the inflationary phase proceeds, the universe tends to the strong coupling phase of the traceless mode. 
At some instant the phase transition occurs, and the correlation becomes short-range 
of order of $1/\Lam_\QG$.
The field fluctuation percolates to localized objects, which may be called ${\it graviball}$ like glueball in QCD, with the size of order $1/\Lam_\QG$.
They eventually decay into the classical matter driving the universe into the Einstein gravity phase.
In order to make the scenario realistic an appropriate field theoretical model is necessary which is able to describe the transition at $\tau_{\Lam}$.

K.H. wishes to thank Y. Yasui for helpful discussions. T.Y. acknowledges N. Sugiyama of NAO/Sokendai for pointing out the ISW effect.


\begin{thebibliography}{99}
\bibitem{cobe}
C. Bennett et al., Astrophys. J. {\bf 464} (1996) L1. 
\bibitem{wmap}
C. Bennett et al., Astrophys. J. Suppl. {\bf 148} (2003) 1. 
\bibitem{guth}
A. Guth, Phys. Rev. {\bf D23} (1981) 347; 
K. Sato, Mon. Not. R. Astron. Soc. {\bf 195} (1981) 467; 
A. Linde, Phys. Lett. {\bf B108} (1982) 389; 
A. Albrecht and P. Steinhardt, Phys. Rev. Lett. {\bf 48} (1982) 1220. 
\bibitem{starobinsky}
A. Starobinsky, Phys. Lett. {\bf B91} (1980) 99; 
A. Vilenkin, Phys. Rev. {\bf D32} (1985) 2511.  
\bibitem{hamada01}
K. Hamada, Mod. Phys. Lett. {\bf A16} (2001) 803. 
\bibitem{sw}
R. Sachs and A. Wolfe, Astrophys. J. {\bf 147} (1967) 73. 
\bibitem{hz}
E. Harrison, Phys. Rev. {\bf D1} (1970) 2726; 
Ya. B. Zel'dovich, Mon. Not. R. Astron. Soc. {\bf 160} (1972) P1.
\bibitem{aw}
L. Abbott and M. Wise, Astrophys. J. {\bf 282} (1984) L47; 
R. Scaramella and N. Vittorio, Astrophys. J. {\bf 353} (1990) 372.
\bibitem{hamada02}
K. Hamada, Prog. Theor. Phys. {\bf 108} (2002) 399.   
\bibitem{hey}
S. Horata, H. Egawa and T. Yukawa, 
Nucl. Phys. {\bf B (Proc. Suppl.) 106} (2002) 971; hep-lat/0209004.  
\bibitem{hh}
K. Hamada and S. Horata, hep-th/0307008. 
\bibitem{kpz}
V. Knizhnik, A. Polyakov and A. Zamolodchikov, Mod. Phys. Lett. {\bf A3} (1988) 819;  
J. Distler and H. Kawai, Nucl. Phys. {\bf B321} (1989) 509; 
F. David, Mod. Phys. Lett. {\bf A3} (1988) 1651.  
\bibitem{riegert}
R. Riegert, Phys. Lett. {\bf 134B} (1984) 56.
\bibitem{am}  
I. Antoniadis and E. Mottola, Phys. Rev. {\bf D45} (1992) 2013.  
\bibitem{amm}
I. Antoniadis, P. Mazur and E. Mottola, 
Phys. Rev. Lett. {\bf 79} (1997) 14. 
\bibitem{hu-sugiyama}
W. Hu and N. Sugiyama, Phys. Rev. {\bf D51} (1995) 2599.
\bibitem{ll}
A. Liddle and D. Lyth, {\it Cosmological Inflation and Large-Scale Structure}, 
Cambridge Univ. Press, 2000.
\end{thebibliography}
\end{document}